# Transient Simulation of Grid-Feeding Converter System for Stability Studies Using Frequency Response Optimized Integrators

Sheng Lei[1,2], Student Member, IEEE, and Alexander Flueck[1], Senior Member, IEEE
[1]Department of Electrical and Computer Engineering
Illinois Institute of Technology
Chicago, IL, USA
[2]Mathematics and Computer Science Division
Argonne National Laboratory
Lemont, IL, USA
Email: slei3@hawk.iit.edu, and flueck@iit.edu

*Abstract*—**A grid-feeding converter system is added to a novel power system transient simulation scheme based on frequency response optimized integrators considering second order derivative. The converter system and its implementation in the simulation scheme are detailed. Case studies verify the accuracy and efficiency of the simulation scheme. Furthermore, this paper proposes and justifies extending the simulation scheme by integrating commonly used numerical integrators considering first order derivative for part of the studied system. The proposed extension has an insignificant impact on the accuracy of the simulation scheme while significantly enhancing its efficiency. It also reduces the development burden in adding new devices.**

## I. Introduction

Increasing research attention is being paid to stability analysis of unbalanced distribution systems or microgrids [1]-[2]. Electromagnetic transient (EMT) simulation has been applied in these studies due to its capability to model unbalanced devices and systems in detail [1], [3]. Unfortunately traditional EMT simulation is computationally inefficient because small step sizes have to be adopted. Several modified EMT simulation schemes have been proposed to enable large step sizes so that computational efficiency is improved [4]-[5]. In [5], the authors put forward a novel transient simulation scheme based on frequency response optimized integrators considering second order derivative.

Grid-connected converter systems, including the converters and the related controllers, can be classified into grid-feeding, grid-forming and grid-supporting ones according to their modes of operation [6]. Among them, grid-feeding converter systems have the largest share [6]. It is thus important to study the dynamics and the impacts on system stability of this type of device.

In this paper, a typical grid-feeding converter system is added to the novel transient simulation scheme [5]. Furthermore, this paper explores the possibility of extending the simulation scheme by integrating commonly used numerical integrators considering first order derivative. The main contributions of this paper are twofold. First, a grid-feeding converter system is included as a new type of device to enhance the functionality of the simulation scheme. The components and the implementation of the converter system in the simulation scheme are introduced in detail. Accuracy and efficiency of the simulation scheme applied to the converter system are verified. Second, extension of the simulation scheme by integrating the commonly used implicit trapezoidal method and backward Euler method is proposed. The feasibility of the proposed extension is justified in that the efficiency of the simulation scheme is enhanced while the accuracy is basically not undermined.

The remainder of this paper is organized as follows. Section II reviews the state of the art of frequency response optimized integrators and the novel transient simulation scheme along with introduction of the proposed extension. Section III introduces the converter system under study and how the numerical integrators are applied. Section IV verifies the accuracy and efficiency of the simulation scheme and its extension applied to the converter system via case studies. Section V concludes the paper and points out some directions for future research.

## II. State of the Art

### A. Frequency Response Optimized Integrators

Consider a general ordinary differential equation (ODE)
$$\dot{x} = f(t,x,u) \qquad (1)$$
where $x$ is the state variable; $t$ is the time instant; $u$ is the input; $f$ is a function depending on $t$, $x$ and $u$. (1) may be discretized and solved by a numerical integrator considering first order derivative as (2)
$$x_t = x_{t-h} + b_0 \dot{x}_t + b_{-1} \dot{x}_{t-h} \qquad (2)$$

where $h$ is the step size; $b_0$ and $b_{-1}$ are coefficients to be determined. A specific selection for these coefficients determines a numerical integrator. For example, the backward Euler method [7] has $b_0 = h$ and $b_{-1} = 0$. The implicit trapezoidal method [7] has $b_0 = h/2$ and $b_{-1} = h/2$.

ODE (1) may be discretized by a numerical integrator considering second order derivative as

$$x_t = x_{t-h} + b_0 \dot{x}_t + b_{-1} \dot{x}_{t-h} + c_0 \ddot{x}_t + c_{-1} \ddot{x}_{t-h} \quad (3)$$

where $b_0$, $b_{-1}$, $c_0$ and $c_{-1}$ are coefficients. Again a specific selection for the coefficients determines a numerical integrator. In (3), the second order derivative of the state variable is calculated as

$$\ddot{x} = \frac{\partial f}{\partial t} + \frac{\partial f}{\partial x}\dot{x} + \frac{\partial f}{\partial u}\dot{u} \quad (4)$$

The derivative of the input is required in (4). It may be externally given or numerically calculated according to the input.

The coefficients in (3) can be selected so that the numerical integrator introduces no error at a specified nonzero angular frequency $\omega_{select}$ despite the specific value of the step size. Note that the state variable of an ODE may be dominated by a nonzero frequency. For instance, the voltage and current waveforms in a power system are dominated by the utility frequency (50 or 60 Hz). If this is the case, $\omega_{select}$ can be specified at the dominant frequency so that the overall error introduced by the discretization process is significantly reduced. Consequently large step sizes can be directly used to shorten time consumption without sacrificing accuracy. From the numerical error viewpoint, the frequency response of the numerical integrator is thus optimized. Similarly those numerical integrators which are accurate for slow variants can also be considered as frequency response optimized if they are applied to the corresponding ODEs. In fact, the implicit trapezoidal method and the backward Euler method are accurate for slow variants [4].

Four frequency response optimized integrators considering second order derivative are introduced in [5], namely Integrators A-D. They are listed in Table I. Integrators A and B are accurate for state variables with a nonzero dominant frequency component at $\omega_{select}$ and 0 Hz. Integrators C and D are accurate for slow variants. Integrator A is more accurate than Integrator B while Integrator C is more accurate than Integrator D. Detailed discussion of these numerical integrators can be found in [5].

TABLE I. FREQUENCY RESPONSE OPTIMIZED INTEGRATORS CONSIDERING SECOND ORDER DERIVATIVE

| Coefficient | Numerical Integrator | | | |
|---|---|---|---|---|
| | A | B | C | D |
| $b_0$ | $\dfrac{h}{2}$ | $\dfrac{sin(\omega_{select}h)}{\omega_{select}}$ | $\dfrac{h}{2}$ | $h$ |
| $b_{-1}$ | $\dfrac{h}{2}$ | 0 | $\dfrac{h}{2}$ | 0 |
| $c_0$ | $\dfrac{-1}{\omega_{select}^2} + \dfrac{h}{2\omega_{select}}cot\left(\dfrac{\omega_{select}h}{2}\right)$ | $\dfrac{cos(\omega_{select}h)-1}{\omega_{select}^2}$ | $-\dfrac{h^2}{12}$ | $-\dfrac{h^2}{2}$ |
| $c_{-1}$ | $\dfrac{1}{\omega_{select}^2} - \dfrac{h}{2\omega_{select}}cot\left(\dfrac{\omega_{select}h}{2}\right)$ | 0 | $\dfrac{h^2}{12}$ | 0 |

## B. The Novel Transient Simulation Scheme and Its Extension

A power system is typically modeled as a set of differential-algebraic equations (DAEs) [3]. Given system parameters and initial conditions, a transient simulation scheme numerically solves the equation set. The ODEs involved are discretized into algebraic ones by numerical integrators so as to be solved numerically. The implicit trapezoidal method is commonly used in power system transient simulation [3].

The novel transient simulation scheme is based on frequency response optimized integrators considering second order derivative [5]. Instead of using a unique numerical integrator for the entire system, multiple numerical integrators are selected according to the frequency spectrum of the state variable of individual ODEs. Integrator A is used for the ODEs with state variables that are dominated by the utility frequency while Integrator C is used for ODEs with state variables that are slowly varying.

As commonly adopted in power system transient simulation, the simulation scheme adopts a fixed step size during a simulation run. The equations of power system network and other devices are solved simultaneously via iteration at each time step, which ensures consistency of the whole system so that high fidelity is achieved. Discontinuities are dealt with by a technique similar to the Critical Damping Adjustment (CDA) [8]. Immediately after a discontinuity, Integrators A and C are temporarily replaced by Integrators B and D respectively.

In this paper, the possibility of extending the simulation scheme, which is initially purely based on numerical integrators considering second order derivative, by including numerical integrators considering first order derivative is explored. The implicit trapezoidal method is applied to the part of the system where state variables vary slowly, if at all. When a discontinuity is encountered, the backward Euler method temporarily replaces the implicit trapezoidal method.

## III. GRID-FEEDING CONVERTER SYSTEM

### A. Overview of Grid-Feeding Converter System

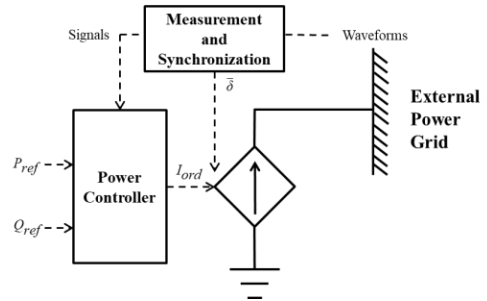

Figure 1. Grid-feeding converter system.

A diagram of the grid-feeding converter system studied in this paper is shown in Fig. 1. The structure is based on [6]. The converter system aims at regulating its real and reactive power outputs. Real and reactive power references are sent to the power controller, which accordingly generates a control order for the remaining part of the converter system. The overall behavior of the remaining part can be understood as an equivalent controlled AC current source. The control order from the power controller is thus a current order, which is typically expressed as a phasor in the device reference frame. The cumulative phase angle of the terminal voltage of the converter system is required to provide information about the system reference frame, so that the equivalent controlled AC current source is able to feed the correct current waveforms into the external power grid.

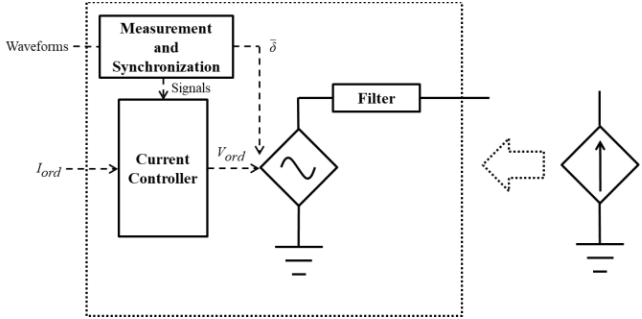

Figure 2. Equivalent controlled AC current source.

The components of the equivalent controlled AC current source are shown in Fig. 2, including a filter, a voltage source converter (VSC) and related controllers. The current order phasor is sent to the current controller, which generates a voltage order phasor in the device reference frame. Receiving the voltage order phasor and the cumulative phase angle of the terminal voltage, the VSC provides the required voltage waveforms. The output waveforms further go through a filter so that the currents fed into the external power grid approximate sinusoids at the utility frequency.

A measurement and synchronization mechanism measures the voltage and current waveforms and converts them into phasors used by other controllers. The mechanism also measures the cumulative phase angle of the terminal voltage, a process which is known as grid synchronization [6]. Moreover, it measures real and reactive power outputs of the converter system.

### B. L Filter and VSC

The filter and VSC introduced previously form the power part of the converter system. The filter for a grid-feeding converter system is usually an inductor (L filter) [6], [9]. In this paper the inductance of each phase is assumed to be the same and denoted by $L_f$. The current waveforms through the L filter are dominated by the frequency component at the utility frequency. Therefore Integrator A is used for the filter.

In power system stability studies, a VSC is typically simplified as a controlled AC voltage source [2], [10]. Its DC side is not further modeled. It receives a voltage order and accordingly generates three-phase balanced voltage waveforms in positive sequence. As mentioned before, voltage order phasor in the device reference frame and the cumulative phase angle of the terminal voltage is passed to the VSC. The corresponding Phase A voltage waveform of the VSC is

$$v_A = V_{ord,m} \cos(\bar{\delta} + V_{ord,a,dev}) \quad (5)$$

where $V_{ord,m}$ is the voltage order phasor magnitude; $V_{ord,a,dev}$ is the voltage order phasor angle in the device reference frame; $\bar{\delta}$ is the cumulative phase angle of the terminal voltage. Acquisition of $\bar{\delta}$ will be detailed later in this section. The VSC is a pure algebraic component in that no differential equation is involved. Hence it needs no numerical integrator.

### C. Power Controller

A design of the power controller [9] is shown in Fig. 3. It receives externally given real and reactive power references and generates current order phasor for the current controller.

The signals inside the power controller are basically slow variants. In this paper, two computational schemes are considered for the power controller. In Scheme 1, Integrator C is used. In Scheme 2, the implicit trapezoidal method is used.

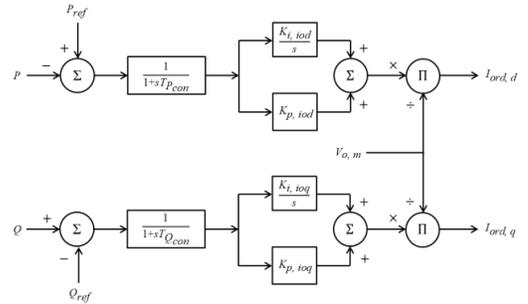

Figure 3. Power controller.

### D. Current Controller

A design of the current controller [6], [9], [11] is shown in Fig. 4. It receives a current order phasor from the power controller and generates a voltage order phasor for the VSC. Integrator C is used for the current controller.

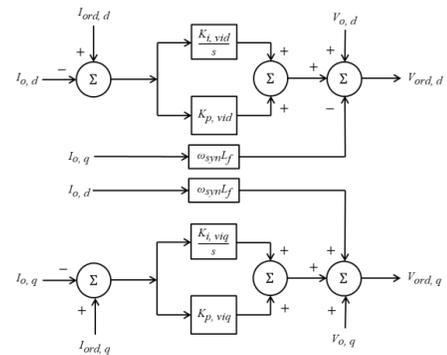

Figure 4. Current controller.

### E. Measurement and Synchronization Mechanism

*1) In-phase and quadrature signal generation:* Applying Clarke's transformation to the three-phase components of a quantity $x$, which can be the terminal voltage or the current through the L filter, the in-phase and quadrature signals are calculated as follows

$$\begin{pmatrix} x_{in} \\ x_{qu} \end{pmatrix} = \frac{2}{3} \begin{pmatrix} \cos(0) & \cos(-\frac{2}{3}\pi) & \cos(\frac{2}{3}\pi) \\ -\sin(0) & -\sin(-\frac{2}{3}\pi) & -\sin(\frac{2}{3}\pi) \end{pmatrix} \begin{pmatrix} x_A \\ x_B \\ x_C \end{pmatrix} \quad (6)$$

After some algebraic operations, it can be verified that if the system is three-phase balanced while the waveforms are purely sinusoidal, the in-phase signal is equal to the Phase A component; the quadrature signal is lagging the in-phase signal by π/2. This is a pure algebraic component so no numerical integrator is required.

*2) Phase shift:* A phase shift factor is applied to the in-phase and quadrature signals to calculate the real and imaginary parts of the corresponding phasor in the device reference frame

$$X = X_d + jX_q = e^{-j\bar{\delta}}(x_{in} + jx_{qu}) \quad (7)$$

This is also a pure algebraic component so no numerical integrator is needed.

*3) Phase-locked loop (PLL):* A PLL [6] is shown in Fig. 5. The imaginary part of the terminal voltage phasor is sent to the PLL as the input. The output is the cumulative phase angle of the terminal voltage. In transient simulation, as the absolute time is available, an alternative implementation may be applied, in which the phase angle regarding the synchronously rotating referece frame is calculated instead, which is also understood as the phasor angle. The cumulative phase angle is then calculated as

$$\bar{\delta} = \omega_{syn} t + \delta \quad (8)$$

where $\omega_{syn}$ is the synchronous angular frequency; $\delta$ is the phasor angle. Integrator C is used for the PLL.

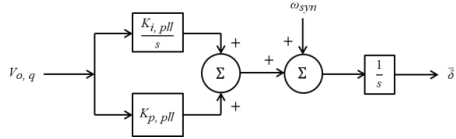

Figure 5. PLL.

*4) Real power output, reactive power output and terminal voltage phasor magnitude measurements:* The real power output, reactive power output and terminal voltage phasor magnitude are first pre-calculated as [11]

$$P_{pre} = V_{o,d}I_{o,d} + V_{o,q}I_{o,q}, \quad Q_{pre} = -V_{o,d}I_{o,q} + V_{o,q}I_{o,d}, \quad V_{o,m,pre} = \sqrt{V_{o,d}^2 + V_{o,q}^2} \quad (9)$$

The measured values are then obtained by passing these pre-calculated values to low-pass filters [11]

$$\dot{P} = \frac{1}{T_P}(-P + P_{pre}), \quad \dot{Q} = \frac{1}{T_Q}(-Q + Q_{pre}), \quad \dot{V}_{o,m} = \frac{1}{T_V}(-V_{o,m} + V_{o,m,pre}) \quad (10)$$

where $T_P$, $T_Q$ and $T_V$ are time constants of the low-pass filters. Integrator C is used for these measurements.

## IV. CASE STUDIES

### A. Description of the Test System

The grid-feeding converter system has been added to the MATLAB implementation of the novel transient simulation scheme [5]. The simulation scheme has also been extended as mentioned in Subsection III C. To study the accuracy and efficiency, Scheme 1 (the original simulation scheme) and Scheme 2 (the extension) are applied to a test system; simulation results and time consumption are compared to those obtained from an iterative EMT simulator [5] applied to the same system. Implemented with MATLAB, the iterative EMT simulator has the same structure and execution as the novel transient simulation scheme, except that the commonly used implicit trapezoidal method is adopted for the whole system. Similar implementations of the simulation schemes make it fair to compare their time consumption.

The test system has two busses. A branch connects the two busses; the per-phase impedance is $0.1038 + j0.8416$. A grid-feeding converter system is connected to Bus 1, the real and reactive power references are 1.0 and 0.35 respectively. Dynamic parameters of the converter system are listed in Table II. Constant impedance loads are connected to Bus 1. Phase A, B and C real power loads are 0.24, 0.3 and 0.36; Phase A, B and C reactive power loads are 0.072, 0.09 and 0.108. Ideal AC voltage sources are connected to Bus 2. Phase A, B and C voltage phasor magnitudes are 1.03, 1.0 and 0.98; Phase A, B and C voltage phasor angles are 0.0°, -121.0° and 118.0°.

TABLE II. DYNAMIC PARAMETERS OF THE CONVERTER SYSTEM

| L Filter | | Power Controller | | Measurement and Synchronization Mechanism | |
|---|---|---|---|---|---|
| $L_f$ | $4.2441 \times 10^{-4}$ | $T_{P_{con}}$ | 0.1 | | |
| **Current Controller** | | $K_{i, iod}$ | 10 | $K_{i, pll}$ | 9000 |
| $K_{i, vid}$ | 0.01 | $K_{p, iod}$ | 1.3 | $K_{p, pll}$ | 150 |
| $K_{p, vid}$ | 0.25 | $T_{Q_{con}}$ | 0.1 | $T_P$ | 0.02 |
| $K_{i, viq}$ | 0.01 | $K_{i, ioq}$ | 10 | $T_Q$ | 0.02 |
| $K_{p, viq}$ | 0.25 | $K_{p, ioq}$ | 1.3 | $T_V$ | 0.02 |

Refer to Section III for meanings of the parameters.

Simulation runs start at 0.0 s. At 0.2 s, a Phase-B-to-Ground fault is applied at Bus 1 with a fault resistance of 0.1 p.u.. At 0.4 s, the fault is cleared. The simulation duration is 2.0 s.

### B. Results

Results from the iterative EMT simulator with a tiny step size of 5 μs are used as the reference. Fig. 6 shows the real and reactive power outputs of the converter system. The converter system succeeds in regulating its power outputs according to the references. Results from Scheme 1, Scheme 2 and the iterative EMT simulator with larger step sizes are compared to the reference in Fig. 7. For better visualization, only the comparison regarding the real power output around 0.46 s is presented. Both Scheme 1 and 2 are more accurate than the iterative EMT simulator with the same or one-half the step size.

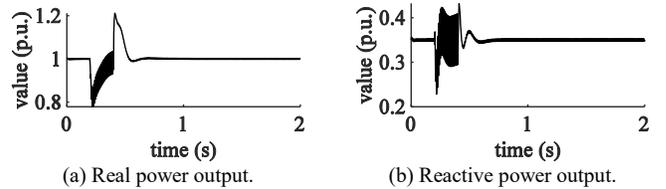

(a) Real power output. (b) Reactive power output.

Figure 6. Power outputs of the converter system.

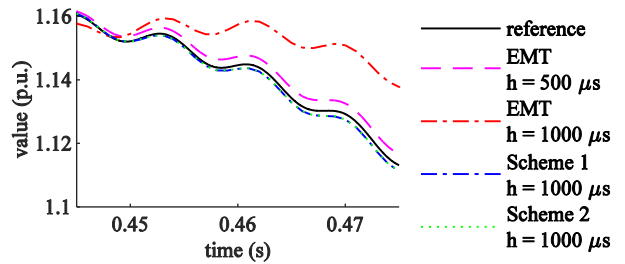

Figure 7. Comparison to the reference.

To quantitatively study the accuracy of the simulation schemes, the following error measurement is used. Suppose that $x$ is a variable to be studied. The relative error regarding $x$ of a simulation scheme given a step size is

$$err(x) = \|x_{com} - x_{ref}\|_2 / \|x_{ref}\|_2 \times 100 \quad (11)$$

where $x_{com}$ is the computed value; $x_{ref}$ is the reference value. The 2-norm is calculated at common time instants of $x_{com}$ and $x_{ref}$.

This paper considers the averaged relative error of Bus 1 voltage waveforms and the relative error of the measured Bus 1 voltage phasor angle. These two errors are referred to as voltage error and phase error respectively hereafter for simplicity. Voltage error is defined as

$$ERR(v) = \frac{1}{3} \sum_{i \in \{Phase\ A,\ B,\ C\}} err(v_i) \quad (12)$$

where $v_i$ is the voltage at a node of Bus 1. Voltage error and phase error of Schemes 1 and 2 and the iterative EMT simulator are listed in Table III. According to voltage error, Schemes 1 and 2 are more than 4 times more accurate than the iterative EMT simulator; according to phase error, the accuracy is more than 10 times better. Schemes 1 and 2 are more robust than the iterative EMT simulator in that the iterative EMT simulator diverges with large step sizes while Schemes 1 and 2 still converge. It is interesting to note that Scheme 2 becomes even more accurate than Scheme 1 as the step sizes go larger; the accuracy of Schemes 1 and 2 deteriorates dramatically when the step size turns from 2000 to 4000 μs.

To study their efficiency, the simulation schemes are applied to an enlarged test system which consists of 10 copies of the one introduced in Subsection A. Here only sequential computation is performed; no parallel computing is considered. Time consumption of the simulation schemes with different step sizes is listed in Table IV. If the same step size is used, Schemes 1 and 2 spend around twice the time consumption of the iterative EMT simulator; Scheme 2 always consumes less time than Scheme 1.

TABLE III. ACCURACY COMPARISON OF THE SIMULATION SCHEMES

| Step Size (μs) | Voltage Error | | | Phase Error | | |
|---|---|---|---|---|---|---|
| | Scheme 1 | Scheme 2 | EMT | Scheme 1 | Scheme 2 | EMT |
| 125 | 0.0263 | 0.0263 | 0.1582 | 0.0004 | 0.0006 | 0.0245 |
| 250 | 0.0654 | 0.0655 | 0.3141 | 0.0036 | 0.0041 | 0.0999 |
| 500 | 0.0688 | 0.0689 | 0.5729 | 0.0261 | 0.0264 | 0.4263 |
| 1000 | 0.1860 | 0.1778 | 2.4619 | 0.1280 | 0.1183 | 2.3736 |
| 2000[a] | 0.5263 | 0.5013 | -- | 0.4312 | 0.4157 | -- |
| 4000[a] | 15.4370 | 3.0339 | -- | 15.5255 | 3.8366 | -- |

a. The iterative EMT simulator diverges with this step size.

TABLE IV. EFFICIENCY COMPARISON OF THE SIMULATION SCHEMES

| Step Size (μs) | Time Consumption | | |
|---|---|---|---|
| | Scheme 1 | Scheme 2 | EMT |
| 125 | 958.56 | 912.87 | 474.09 |
| 250 | 492.31 | 463.92 | 239.57 |
| 500 | 317.12 | 301.67 | 151.28 |
| 1000 | 168.79 | 158.43 | 82.39 |
| 2000[a] | 85.67 | 81.15 | -- |
| 4000[a] | 45.77 | 44.21 | -- |

a. The iterative EMT simulator diverges with this step size.

*C. Comments*

Schemes 1 and 2 can use large step sizes to shorten the time consumption for simulation runs while maintaining satisfactory accuracy. Based on the many experiments that the authors have performed, a step size of 0.5 to 2 ms is suggested for stability studies on unbalanced power systems with high penetration of distributed energy resources (DERs), as a good trade-off between efficiency and accuracy.

The proposed extension indeed enhances the efficiency. An interesting discovery of this paper is that the proposed extension even increases the accuracy as well when millisecond-level step sizes are used. The generality of this phenomenon deserves more investigation. The extension of the novel transient simulation scheme is appealing and promising. One extra advantage is existing routines of some devices for traditional EMT simulation can be directly merged, which will save a lot of development burden.

V. CONCLUSION AND FUTURE WORK

A grid-feeding converter system has been included into the novel transient simulation scheme. Extension of the simulation scheme by integrating commonly used numerical integrators is proposed. Accuracy and efficiency of the simulation scheme and its extension are validated as applied to the converter system. Feasibility of the proposed extension is justified.

In the future, the simulation scheme and its extension may be applied to other devices and systems to further study the numerical error properties and the acceleration effects.

REFERENCES

[1] M. Farrokhabadi et al., "Microgrid Stability Definitions, Analysis, and Examples," in IEEE Transactions on Power Systems, vol. 35, no. 1, pp. 13-29, Jan. 2020.
[2] W. Du et al., "Modeling of Grid-Forming and Grid-Following Inverters for Dynamic Simulation of Large-Scale Distribution Systems," in IEEE Transactions on Power Delivery.
[3] H. W. Dommel, EMTP Theory Book. Vancouver, BC, Canada: Microtran Power System Analysis Corporation, May 1992.
[4] P. Zhang, J. R. Marti and H. W. Dommel, "Shifted-Frequency Analysis for EMTP Simulation of Power-System Dynamics," in IEEE Transactions on Circuits and Systems I: Regular Papers, vol. 57, no. 9, pp. 2564-2574, Sept. 2010.
[5] S. Lei and A. Flueck, "Efficient Power System Transient Simulation Based on Frequency Response Optimized Integrators Considering Second Order Derivative," presented at the 2020 IEEE PES General Meeting. [Online]. Available: https://arxiv.org/abs/2005.00964
[6] J. Rocabert, A. Luna, F. Blaabjerg and P. Rodríguez, "Control of Power Converters in AC Microgrids," in IEEE Transactions on Power Electronics, vol. 27, no. 11, pp. 4734-4749, Nov. 2012.
[7] R. M. Corless and N. Fillion, A Graduate Introduction to Numerical Methods. New York, NY, USA: Springer-Verlag, 2013.
[8] J. R. Marti and J. Lin, "Suppression of numerical oscillations in the EMTP power systems," in IEEE Transactions on Power Systems, vol. 4, no. 2, pp. 739-747, May 1989.
[9] E. Muljadi, M. Singh and V. Gevorgian, "User guide for PV dynamic model simulation written on PSCAD platform", Nov. 2014.
[10] A. A. van der Meer, M. Gibescu, M. A. M. M. van der Meijden, W. L. Kling and J. A. Ferreira, "Advanced Hybrid Transient Stability and EMT Simulation for VSC-HVDC Systems," in IEEE Transactions on Power Delivery, vol. 30, no. 3, pp. 1057-1066, June 2015.
[11] N. Pogaku, M. Prodanovic and T. C. Green, "Modeling, Analysis and Testing of Autonomous Operation of an Inverter-Based Microgrid," in IEEE Transactions on Power Electronics, vol. 22, no. 2, pp. 613-625, March 2007.